\documentclass[12pt]{iopart}
\usepackage{mathrsfs}
\usepackage{iopams}
\usepackage{bbding}
\usepackage{bm}
\begin{document}

\title{The decoherence dynamics of the multipartite entanglement in non-Markovian environment}

\author{Zhi He, Jian Zou, Bin Shao and Shu-Yan Kong}
\address{Department of Physics, School of Science, Beijing Institute of Technology, Beijing 100081, People's Republic of China}
\ead{zoujian@bit.edu.cn}

\begin{abstract}
We consider four two-level atoms interacting with independent
non-Markovian reservoirs with detuning. We mainly investigate the
effects of the detuning and the length of the reservoir
correlation time on the decoherence dynamics of the multipartite
entanglement. We find that the time evolution of the entanglement
of atomic and reservoir subsystems is determined by a parameter,
which is a function of the detuning and the reservoir correlation
time. We also find that the decay and revival of the entanglement
of the atomic and reservoir subsystems are closely related to the
sign of the decay rate. We also show that the cluster state is the
most robust to decoherence comparing with Dicke, GHZ, and W states
for this decoherence channel.
\end{abstract}
\maketitle

\section{Introduction}

Entanglement plays a central role in quantum information
processing. Since the maintenance of entangled states is very
important in quantum information processing systems, the study of
the effect of decoherence on entangled states would be of
considerable importance from theoretical as well as experimental
point of view. Recently entanglement of open quantum systems has
attracted considerable attention due to its significance for both
fundamentals and applications of quantum information processing
\cite{Ficek2006, Dodd2004, Yu2004,
Sun2007,Lopez2008,Qasimi2008,Yu2009}. Yu and Eberly \cite{Yu2004}
 found that the decay of a single qubit
coherence can be slower than the decay of entanglement, and they
named the abrupt disappearance of entanglement at a finite time
entanglement sudden death (ESD), which has been experimentally
tested very recently \cite{Almeida2007}. Later L$\rm\acute{o}$pez
\it et al. \rm \cite{Lopez2008} revealed that when the bipartite
entanglement suddenly disappears, the entanglement of
corresponding reservoir suddenly and necessarily appears, which is
called entanglement sudden birth (ESB). Although the bipartite
entanglement is well understood in many aspects, until now the
multipartite entanglement is far from clear and thus deserves
profound understandings. Several important applications of
multipartite entangled states, like quantum error correction,
quantum computing, \it etc. \rm have been found (for recent
reviews see Refs.\cite{Horodecki0702,Amico2008,Perez-Garcia2007}
and references therein). Recently, people investigated the
decoherence dynamics of multipartite entangled states, for
instance, G$\rm{\ddot{u}}$hne, Bodoky and Blaauboer
\cite{Gohne2008} discussed some multipartite entanglement
properties under the influence of a global dephasing process using
the geometric measure of entanglement, and they showed that the
Dicke state is the most robust to decoherence comparing with GHZ,
cluster, and W states. Borras and Majtey \it et al. \rm
\cite{Borras2009} investigated the robustness of highly entangled
multiqubit states under different decoherence channels
\cite{Nielsen2000}, and later, they further studied the geometry
of robust entangled multiparticle states under decoherence
\cite{Majtey0904}.

In realistic physical systems the assumption of Markovian dynamics
can only be an approximation. Generally speaking the quantum
system of interest interacts with the reservoirs, whose spectral
density strongly varies with frequency, which is called
non-Markovian open quantum systems \cite{Breuer2002}.
Non-Markovian dynamics is characterized by the existence of
 a memory time scale during which information that has been transferred from the system to the environment can
 flow back into the system.  The non-Markovian systems appear in many
branches of physics, such as quantum optics
\cite{Breuer2002,Lambropoulos2000,Gardiner1999}, solid state
physics \cite{Lai2006}, and quantum chemistry \cite{Shao2004}. In
quantum information processing, the non-Markovian character of
decoherence has to be considered \cite{Aharonov2006}. The
non-Markovian dynamics has drawn great attentions including
continuous-variable
\cite{Mancini2005,Ban2006,Maniscalco2007,Liu2007,An2007,Shiokawa2009,Vasile2009}
and discrete-variable systems
 \cite{Breuer2004,Piilo2008,Breuer2009,Mazzola2009,Bellomo2007,Zhang2009,Zhang2010,Zhou2009,Hamadou-Ibrahim12010}.
Very recently $N$ qubits initially in the mixed GHZ-type  and
W-type states interacting with independent structured reservoirs
have been investigated and it is found that the $N$-qubit
entanglement revivals are related to the qubit number $N$ and the
purity of the initial state of $N$ qubits \cite{Zhang2010}. The
measure of the degree of non-Markovian behavior
\cite{Breuer-Laine2009} and a basic relation between the quantum
 Fisher information  flow and non-Markovianity have been proposed for quantum dynamics of open systems
 \cite{lu}.

In this paper, we are very interested in the decoherence dynamics
of multipartite entanglement under the  influence of non-Markovian
reservoir. We mainly investigate the effects of the detuning and
the length of the reservoir correlation time on the decoherence
dynamics of the multipartite entanglement. We consider that four
noninteracting two-level atoms interact with four  independent
non-Markovian reservoirs with detuning. We let the four atoms
initially be prepared in different four particle entangled states
(cluster, Dicke, GHZ, and W states). We find that the dynamical
behaviors of the entanglement of the atomic and reservoir
subsystems for various initial states are determined by a complex
parameter, which is a function of the detuning and the reservoir
correlation time. The real part of this parameter is closely
related to the decay of the entanglement and the imaginary part of
it is closely related the oscillations of the revivals, and all
the dynamical behavior of the entanglement can be uniformly
explained by this parameter. We also find a close relationship
between the decay rate and the entanglement of the atomic and
reservoir subsystems, i.e., the decay or revival of the
entanglement of the atomic and reservoir subsystems are decided by
the sign of the decay rate. We also find that the cluster state is
the most robust to decoherence, which is different from the
results of Ref.\cite{Gohne2008}. Here we consider the amplitude
damping process, while Ref.\cite{Gohne2008} discuss the global
dephasing process. The paper is organized as follows: In Sec. 2,
we present the model which consists of four atoms interacting with
individual non-Markovian reservoirs, and introduce the
multipartite entanglement measure. We present and analyze the
results in Sec. 3.  Finally, we give some conclusions in Sec. 4.

\section{Model}

We consider a system of four identical noninteracting two-level
atoms, each of them coupled to its own reservoir. For simplicity,
we assume that each corresponding reservoir is initially in the
vacuum state. Due to the independence of each atom, we only need
to discuss the problem of a single atom interacting with its
corresponding reservoir. The Hamiltonian of the interaction
between a single atom and $N$-mode reservoir under the
rotating-wave approximation can be written as ($\hbar=1$),
\begin{equation}
\hat{H}=\omega_a\hat{\sigma}_+\hat{\sigma}_-+\sum_{k=1}^N\omega_k\hat{a}_k^\dagger
\hat{a}_k+\sum_{k=1}^Ng_k(\hat{\sigma}_-\hat{a}_k^\dagger
+\hat{\sigma}_+\hat{a}_k),
\end{equation}
where $\hat{\sigma}_+=|1\rangle\langle0|$ and
$\hat{\sigma}_-=|0\rangle\langle1|$, are the Pauli raising and
lowering operators for the atom respectively, and $\omega_a$ is
the transition frequency of each atom. $\hat{a}_k^\dagger$ and
$\hat{a}_k$ are the creation and annihilation operators with
frequency $\omega_k$ for the reservoir mode $k$, and $g_k$ is the
corresponding coupling constant.

Here we consider one excitation case, namely, the atom and
corresponding reservoir are initially in the excited state and
vacuum state respectively, i.e.,
$|\psi_0\rangle=|1\rangle_{a}\otimes|\bar{0}\rangle_r$, where
$|\bar{0}\rangle_r=\prod_{k=1}^N|0_k\rangle_r$. The subscripts $a
$ and $r$ refer to the atom and the corresponding reservoir
respectively. So the state of the total system at any time $t$ can
be denoted by
\begin{equation}
|\psi_t\rangle=\nu(t)|1\rangle_{a}|\bar{0}\rangle_r+\sum_{k=1}^ND_k(t)|0\rangle_{a}|1_k\rangle_r,
\end{equation}
where the state $|1_k\rangle_r$ represents the reservoir having
one excitation in mode $k$.

Similar to the method used in Ref.\cite{Breuer2002}, we can obtain
a closed equation for the coefficient $\nu(t)$ in Eq.(2),
\begin{equation}
\dot{\nu}(t)=-\int\limits_0^tf(t-t_1)\nu(t_1)dt_1,
\end{equation}
where the kernel $f(t - t_1 ) = \int {d\omega J(\omega )} \exp
[i(\omega_a-\omega)(t - t_1 )]$ is related to the spectral density
$J(\omega)$ of the reservoir. We take a Lorentzian spectral
density of the reservoir \cite{Breuer2002,Breuer-Laine2009}
 \begin{equation}
J(\omega ) = \frac{1}{{2\pi }}\frac{{\gamma _0 \lambda ^2
}}{{(\omega_a  - \delta -\omega)^2  + \lambda ^2 }},
\end{equation}
the center of which is detuned from the transition frequency
$\omega_a$ of the two-level atom by an amount $\delta$. And the
parameter $\lambda$ defines the spectral width of the coupling,
which is associated with the reservoir correlation time by the
relation $\tau_B=\lambda^{-1}$ and the parameter $\gamma_0$ is
related to the relaxation time scale $\tau_R$ by the relation
$\tau_R=\gamma_0^{-1}$.

By making the Laplace transformation of Eq.(3), we can obtain the
solution of $\nu(t)$ with initial condition $\nu(0)=1$,
 \begin{equation}
\nu(t) = e^{ - (\lambda  - i\delta )t/2} [\cosh ({{\chi t} \mathord{\left/
 {\vphantom {{\chi t} 2}} \right.
 \kern-\nulldelimiterspace} 2}) + \frac{{\lambda  - i\delta }}{\chi }\sinh ({{\chi t} \mathord{\left/
 {\vphantom {{\chi t} 2}} \right.
 \kern-\nulldelimiterspace} 2})],
\end{equation}
where
\begin{equation}
\chi=\sqrt{(\lambda-i\delta)^2-2\gamma_0\lambda}.
\end{equation}
Furthermore, if we take the form of a collective state of the
reservoir, namely, letting
$|\bar{1}\rangle_r=(1/\mu(t))\sum_{k=1}^ND_k(t)|1\rangle_r$ with
$\mu(t)=\sqrt{1-|\nu(t)|^2}$, Eq.(2) can be rewritten as
\begin{equation}
|\psi_t\rangle=\nu(t)|1\rangle_{a}|\bar{0}\rangle_r+\mu(t)|0\rangle_{a}|1\rangle_r.
\end{equation}

There are many entanglement measures to quantify the bipartite
entanglement such as von Neumann entropy, negativity
\cite{Peres1996,Horodecki1996}, concurrence \cite{Wootters1998},
relative entropy \cite{Vedral1998}, \it etc\rm. However, generally
speaking, for multipartite system the definition of entanglement
measure is difficult. Up to now,  there are some entanglement
measures proposed to quantify the multipartite entanglement, which
include Schmidt measure \cite{Eisert2001}, the geometric measure
of entanglement \cite{Wei2003}, the global entanglement measure
\cite{Meyer2002,Hassan2008}, \it etc\rm. In this paper, we will
use a very popular entanglement measure
\cite{Borras2009,Majtey0904} which is averaged over all possible
bipartitions. The mathematical definition of this measure is
\begin{equation}
E=\frac{1}{[N/2]}\sum_{m=1}^{[N/2]}E^{(m)},
\end{equation}
where
\begin{equation}
E^{(m)}=\frac{1}{N_{\rm bipart}^m}\sum_{i=1}^{N_{\rm
bipart}^m}E(i).
\end{equation}
$E(i)$ indicates the entanglement connected to single bipartition
of the $N$-qubit system, and $E^{(m)}$ denotes the average
entanglement over all nonequivalent bipartitions $N_{\rm
bipart}^m$ between subsets of $m$ qubits and the remaining $N-m$
qubits. If one uses the linear entropy $S_L$ of the reduced
density matrix of the smaller bipartitions to compute $E(i)$, it
will reduce to the well known Meyer-Wallach multipartite
entanglement measure \cite{Meyer2002}. Because in this paper we
will deal with mixed states, we take the negativity to measure the
bipartite entanglement. The normalized negativity is defined as
\cite{Majtey0904}
\begin{equation}
E(i)= \frac{2}{2^m-1}\sum_i|\alpha_i|,
\end{equation}
where $\alpha_i$ is the negative eigenvalue of the partial
transpose matrix for $m$ and the remaining $N-m$ bipartition.

\section{Results and discussions}

We suppose that four two-level atoms are initially prepared in
some famous multipartite entangled states (cluster, Dicke, GHZ,
and W states), and the corresponding reservoirs are initially
prepared in the vacuum state. For W state the initial state of
atom-reservoir system is:
\begin{eqnarray}
\nonumber  |\phi_0\rangle &=&
(|0001\rangle_{a_1a_2a_3a_4}+|0010\rangle_
{a_1a_2a_3a_4}+|0100\rangle_{a_1a_2a_3a_4} \\
  ~&~& +|1000\rangle_{a_1a_2a_3a_4})|\bar{0}\bar{0}
\bar{0}\bar{0}\rangle_{r_1r_2r_3r_4}/2,
\end{eqnarray}
where the subscripts $a_i (i=1,2,3,4)$ represents the atom, and $r_i
(i=1,2,3,4)$ refers to the corresponding reservoir. From Eq.(7), the
evolution of the total system can be obtained
\begin{eqnarray}\label{eq:2}
\nonumber  \left| {\phi _t } \right\rangle&=& [\left|{000}\right\rangle _{a_1 a_2 a_3 }
\left| {\bar 0\bar 0\bar 0}\right\rangle _{r_1 r_2 r_3}(\nu(t)|1\rangle_{a_4}|\bar{0}\rangle_{r_4}+\mu(t)|0\rangle_{a_4}|1\rangle_{r_4})\\
\nonumber ~&~&+ \left| {000}\right\rangle _{a_1 a_2 a_4 } \left|
{\bar 0\bar 0\bar 0}
\right\rangle _{r_1 r_2 r_4 }(\nu(t)|1\rangle_{a_3}|\bar{0}\rangle_{r_3})+\mu(t)|0\rangle_{a_3}|1\rangle_{r_3}) \\
\nonumber  ~ &~& +\left| {000} \right\rangle _{a_1 a_3 a_4 } \left|
{\bar 0\bar 0\bar 0} \right\rangle _{r_1 r_3 r_4 }
   (\nu(t)|1\rangle_{a_2}|\bar{0}\rangle_{r_2})+\mu(t)|0\rangle_{a_2}|1\rangle_{r_2})\\
   ~&~&+\left| {000} \right\rangle _{a_2 a_3 a_4 } \left| {\bar 0\bar 0\bar 0}
   \right\rangle _{r_2 r_3 r_4 } (\nu(t)|1\rangle_{a_1}|\bar{0}\rangle_{r_1}+\mu(t)|0\rangle_{a_1}|1\rangle_{r_1})
   ]/2.
\end{eqnarray}

So the reduced density operator of atomic subsystem $\rho _{a} (t)
= \rm{Tr}\it _{r} (\left| {\phi _t } \right\rangle \left\langle
{\phi _t } \right|)$ and the reduced density operator of reservoir
subsystem $\rho _{r} (t) = \rm{Tr}\it_ {a} (\left| {\phi _t }
\right\rangle \left\langle {\phi _t } \right|)$ can be obtained.
Then from Eq.(8-10) we can obtain the degree of the entanglement of
the atomic and the reservoir subsystems. For convenience, we
denote the degree of the entanglement of the atomic subsystem and
the corresponding reservoir subsystem by $E_a$ and $E_r$,
respectively. Here for W state, $E_a$ and $E_r$ can be obtained,
\begin{equation}
E_a=|[16-16|\nu(t)|^2-(6\sqrt{7|\nu(t)|^4-8|\nu(t)|^2+4}+4\sqrt{2|\nu(t)|^4-2|\nu(t)|^2+1})]|/24,
\end{equation}
and
\begin{equation}
E_r=|[16|\nu(t)|^2-(6\sqrt{7|\nu(t)|^4-6|\nu(t)|^2+3}+4\sqrt{2|\nu(t)|^4-2|\nu(t)|^2+1})]|/24.
\end{equation}
Similarly we can obtain $E_a$ and $E_r$ for the other initial
states, cluster state $\left| {CL_4 } \right\rangle  = (\left|
{0000} \right\rangle  + \left| {0011} \right\rangle  + \left|
{1100} \right\rangle  - \left| {1111} \right\rangle )/2$ , Dicke
state $\left| {D_4 } \right\rangle  = (\left| {0011} \right\rangle
+ \left| {0101} \right\rangle  + \left| {1001} \right\rangle  +
\left| {1100} \right\rangle  + \left| {0110} \right\rangle  +
\left| {1010} \right\rangle )/\sqrt 6 $ , and GHZ state $\left|
{DHZ_4 } \right\rangle  = (\left| {0000} \right\rangle  + \left|
{1111} \right\rangle )/\sqrt 2 $, respectively.

First we consider the resonant case, i.e., $\delta =0$. (i)
$\lambda=10\gamma_0$, which corresponds to the Markovian regime.
In Fig.1 we plot the entanglement evolution of the atomic and
reservoir subsystems for initial cluster, Dicke, GHZ, and W
states. From Fig.1a we can see that the cluster state is the most
robust against decoherence. The result is quite different from
that of Ref.\cite{Gohne2008}, in which the Dicke state is the most
robust against decoherence among GHZ, cluster, and W states. This
is because that the decoherence channel used in
Ref.\cite{Gohne2008} is different from ours. In
Ref.\cite{Gohne2008} the global dephasing channel is considered,
while in this paper we consider the amplitude damping channel.
This means that the most robust multipartite entangled state might
be different for different decoherence channels. It should be
noted that the entanglement measure different from ours is used in
Ref.\cite{Gohne2008}, but we use our entanglement measure to
recalculate the entanglement of Ref.\cite{Gohne2008}, and the
result is the same as that of Ref.\cite{Gohne2008}. From Fig.1 we
can see that the entanglement of the atomic subsystem for all the
initial states decreases monotonically to zero, the entanglement
of the reservoir subsystem for all the initial states increases
monotonically to the steady maximum, and the entanglement
contained initially in the atomic subsystem is finally transferred
into the reservoir subsystem.

(ii) $\lambda=0.1\gamma_0$, which corresponds to the non-Markovian
regime with relatively short reservoir correlation time. In Fig.2
we plot the entanglement evolution of the atomic and reservoir
subsystems for initial cluster, Dicke, GHZ, and W states with
$\lambda=0.1\gamma_0$ and $\delta=0$. The dynamical behaviors of
the entanglement of the atomic and reservoir subsystems in the
non-Markovian regime are quite different from that in the
Markovian regime. It can be seen from Fig.1a and Fig.2a that the
common feature for different initial states in the non-Markovian
regime is that the entanglement of the atomic subsystem decreases
to zero much slowly than that in the Markovian regime, and in the
non-Markovian regime after the entanglement of the atomic
subsystem decays to zero it can revive at later time, which is
quite different from the Markovian case. The reason is that the
information, which the atomic subsystem loses to the reservoir, is
later recovered by the atomic subsystem due to the reservoir
non-Markovian memory. It is noted that in both the Markovian and
non-Markovian  regimes all the initial entanglement of atomic
subsystem $E_a$ will decay and is eventually lost for long times,
and the entanglement of reservoir subsystem $E_r$ gradually
increases to the steady maximum from zero, which can be seen from
Figs.1 and 2. It can be seen from Figs.1b and 2b that in the
non-Markovian regime the entanglement of the reservoir subsystem
$E_r$ at first shows oscillations as a function of time for all
the initial atomic states, and finally the steady maximum
entanglement is achieved, while in the Markovian regime $E_r$
increase to the steady maximum monotonically. All the distinction
between the entanglement properties in the Markovian regime and
that in the non-Markovian regime is induced by the non-Markovian
memory. In other words, in the Markovian regime the information
flow is one directional, namely from atoms to reservoirs, while in
the non-Markovian regime the information flow is bidirectional,
namely the exchange of information back and forth between the
atomic and reservoir subsystems, which causes the oscillations of
the entanglement of the atomic and reservoir subsystems.

(iii) $\lambda=0.01\gamma_0$, which corresponds to the
non-Markovian regime with the relatively long reservoir
correlation time. In Fig.3 we plot the entanglement evolutions of
atomic and reservoir subsystems for the four initial states with
$\lambda=0.01\gamma_0$ and $\delta=0$. Comparing Figs.2a and 3a,
we can find that the revival of the $E_a$ with relatively long
reservoir correlation time is more obvious than that with
relatively short reservoir correlation time, i.e., the amplitude
of revival with relatively long reservoir correlation time is much
larger than that with relatively short reservoir correlation time.
For reservoir subsystem, compared with the case with relatively
short reservoir correlation time, it is more difficult to achieve
the steady maximum of entanglement with relatively long reservoir
correlation time, which can be seen from Figs.2b and 3b. This can
be understood as follows: Increasing the reservoir correlation
time means that the memory effect of the reservoir becomes
stronger, and then the amount of information exchanged between the
atomic and the reservoir subsystems will be enhanced. So the
atomic subsystem can obtain more information from the reservoir
subsystem in the case of relatively long reservoir correlation
time and the revival is stronger, and because of the enhanced
information exchange back and forth the reservoir subsystem will
need more time to achieve the final maximum entanglement.

Now we consider the off-resonant case, i.e., $\delta=8\lambda$.
(i) $\lambda=0.1\gamma_0$. In Fig.4 we plot the entanglement
evolution of atomic and reservoir subsystems for the four initial
states with $\delta=8\lambda$ and $\lambda=0.1\gamma_0$. In the
off-resonant case the entanglement of the atomic subsystem $E_a$
decays to zero with small oscillations, and during each
oscillation $E_a$ can not collapse to zero. And the overall decay
rate is smaller than that in the corresponding resonant case,
which can be seen from Figs.4a and 2a. The entanglement of the
reservoir subsystem $E_r$ at first increases with very small
amplitude oscillations in a very short period of time and then
increases monotonically to the steady entanglement, and the
overall increasing rate is smaller than that in the corresponding
resonant case, which can be seen from Figs.4b and 2b. This can be
easily understood: When the value of the detuning $\delta$
increases, the effective coupling between the atomic and reservoir
subsystem decreases. So the exchange of information between the
atomic subsystem and the reservoir subsystem is not effective and
adequate. (ii) $\lambda=0.01\gamma_0$. In Fig.5 we plot the
entanglement evolution of atomic and reservoir subsystems for the
four initial states with $\delta=8\lambda$ and
$\lambda=0.01\gamma_0$. Comparing Figs.5b and 3b we can find that
due to the increasing of $\delta$, the exchange of information is
not effective, the oscillations of $E_r$ are not adequate, more
specifically $E_r$ can not achieve its maximum during each
oscillation. From Figs.5a and 3a it can be found that increasing
the detuning $\delta$ the period of the revival is shorten, and
the amplitude of the revival increases. This result is very
interesting. As we have mentioned above increasing the detuning
$\delta$ will make the exchange of information less effective,
then why the revival becomes stronger? Now we analyze the
decoherence dynamics of the multipartite entangled states in
detail.

To gain insight in the physical processes characterizing the
decoherence dynamics for different initial states, we consider the
parameter $\chi$, and find that all the above phenomenon can be
uniformly explained by this parameter. From Figs.1-5 it is easy to
find that the dynamical behaviors of the entanglement for
different initial states are very similar. For simplicity in the
following we will take W state as an example. From Eq.(6) we can
see that generally $\chi$ is a complex number, and we will show
that the real part $\mathrm{Re}\chi$ is responsible for the decay
of $E_a$ and the imaginary part $\mathrm{Im}\chi$ is responsible
for the oscillations associated with the revival. From Eq.(13) we
can see that the degree of entanglement is a function of
$|\nu(t)|^2$, which means that all the decoherence dynamics of the
entanglement entirely depends on $\nu(t)$. And it is easy to see
from Eq.(5) that in the long time limit $\nu(t)$ is dominated by
the terms containing the factor
$e^{(-\lambda+|\mathrm{Re}\chi|)t/2}$. From numerical calculations
we find that $|\mathrm{Re}\chi|$ increases with $\delta$ and is
always less than $\lambda$, and $|\mathrm{Im}\chi|$ also increases
with $\delta$. From Eqs.(5) and (13) roughly speaking,
$\lambda-|\mathrm{Re}\chi|$ determines the decay of the
entanglement, which we call it the decay exponent to be
distinguished from the decay rate $\gamma(t)$ \cite{Breuer2002},
and $|\mathrm{Im}\chi|$ determines the basic frequency of the
oscillations in the revivals (it is noted that the overall phase
factor $e^{i\delta t/2}$ in $\nu(t)$ does not make any
contributions to the entanglement). When $\lambda>2\gamma_0$ and
$\delta=0$, from Eq.(6) $\chi$ is a real number, i.e.,
$|\mathrm{Im}\chi|=0$, which corresponds to the Markovian regime.
Hence $E_a$ will decay exponentially to zero without oscillations,
and the revival can not appear. It is easy to prove that when
$\lambda>2\gamma_0$, the decay exponent
$\lambda-\sqrt{\lambda^2-2\gamma_0\lambda}$ is a decreasing
function of $\lambda$, and approaches $\gamma_0$ with the
increasing $\lambda$, the maximum value of which is $2\gamma_0$
occurring at $\lambda=2\gamma_0$. When $\delta= 0$ and
$\lambda<2\gamma_0$, which is corresponding to the non-Markovian
regime, $\chi$ is a pure imaginary, and the oscillations appear.
In this case the decay exponent is just $\lambda$. That is why the
entanglement decay for $\lambda =0.1\gamma_0$ corresponding to the
non-Markovian regime (Fig.2a) is slower than that $\lambda
=10\gamma_0$ corresponding to the Markovian regime (Fig.1a). And
also that is why the entanglement with $\lambda =0.01\gamma_0$ and
$\delta =0$ (Fig.3a) decays much slowly than that with $\lambda
=0.1\gamma_0$ and $\delta =0$ (Fig.2a). Remember that
$|\mathrm{Re}\chi|$ and $|\mathrm{Im}\chi|$ increase with
$\delta$, so the decay exponent $\lambda-|\mathrm{Re}\chi|$
decreases with the increasing of $\delta$. In this way the
envelope of $E_a(t)$ decay more and more slowly with the
increasing of $\delta$, so that during each revival the amplitude
achieved is increasing with the increasing of $\delta$. This
explains why with the increasing the detuning $\delta$ the period
of the revival is shorten, and the amplitude of the revival
increases (see Figs.3a and 5a). Now we consider the dispersive
regime, i.e., $\delta\gg\lambda,\gamma_0$, and in this case in the
long time limit $\nu(t)\approx 1-i\lambda^2/4\delta^2$ and the
steady entanglement of the corresponding atomic subsystem
$E_a\approx [6\sqrt{3}+4+(20+6\sqrt{3})\lambda^4/16\delta^4]/24$
can be achieved. This means that in the dispersive regime the
decay of entanglement $E_a $ is strongly inhibited. We also
calculate the degree of entanglement for the initial five (and
six) particle W state, and we find that the entanglement dynamics
of atomic subsystem for the initial five (or six) particle W state
is almost the same as that of the initial four particle W state,
more specifically the influence of the detuning and the length of
the reservoir correlation time on the dynamical behavior of the
entanglement for the initial five (or six) particle W state is
almost the same as that for the initial four particle W state.
Whenever the degree of entanglement for the four particle case
increases, the degree of entanglement for the corresponding five
(or six) particle case also increases, and whenever the degree of
entanglement for the four particle case decreases, the degree of
entanglement for the five (or six) particle case also decreases.
And the time, at which the entanglement reaches the maximum (or
the minimum), is almost the same  for all the three cases. This
can be understood. Because we find that the entanglement of the
atomic subsystem for five (or six) particle case is also a
function of $|\nu(t)|^2$, which means that the dynamical behavior
of the entanglement for the five (or six) particle is also decided
by the real part and imaginary part of $|\chi|$.

It is well known that in the most general form of a time-local
master equation for the reduced density operator, the decoherence
is induced by the Lindblad (jump) operator with a decay rate
$\gamma(t)$. If the decay rate $\gamma(t)$ is always positive,
this describes the so-called time-dependent Markovian process
\cite{Wonderen2000,Breuer2004,Wolf2008}, but if at least during a
period of time the decay rate $\gamma(t)$ is negative, the
non-Markovian process emerges. Now we also take W state as an
example to show the relation between the decoherence dynamics of
the entanglement and the decay rate $\gamma(t)$. For simplicity we
let $\lambda=0.01\gamma_0$ and $\delta=0$, and in this case the
decay rate $\gamma(t)$ can be expressed as $\gamma(t)=-2\rm
Re\{\dot{\nu(t)}/\nu(t)\}$ \cite{Breuer2002}, where $\nu(t)$ is
obtained by choosing $\delta$=0 in Eq.(5). In Fig.6 we plot $E_a$,
$E_r$ and $\gamma(t)$ as functions of scaled time $\gamma_0t$ for
$\lambda=0.01\gamma_0$ and $\delta=0$. From Fig.6, it is obvious
to see that whenever $\gamma(t)$ (dotted line) takes negative
values, $E_a$ (solid line) begins to revive and increase
monophonically, and the corresponding $E_r$ begins to decrease
monophonically; when $\gamma(t)$ takes positive values, $E_a$ will
begin to decrease monophonically, and $E_r$ begins to increase
monophonically. This can be easily understood: When $\gamma(t)$ is
positive, the information flow is from atomic subsystem to
reservoir subsystem, which means that $E_a$ will decay, and $E_r$
will increase; When $\gamma(t)$ is negative corresponding to the
memory effect of the reservoir, the information flow is from the
reservoir subsystem to atomic subsystem, so $E_a$ will revive and
$E_r$ will decay.

\section{Conclusions}

In this paper, we have considered four atoms with initial
entanglement interact with independent non-Markovian reservoirs.
We have analyzed the decoherence dynamics for various initial
states in Markovian ($\lambda=10\gamma_0$), weak non-Markovian
($\lambda=0.1\gamma_0$) and strong non-Markovian
($\lambda=0.01\gamma_0$) regimes, with and without the detunings.
We have found that the decoherence dynamics of the atomic and
reservoir subsystems strongly depends on a parameter, which is
decided by the detuning and the reservoir correlation time, and
all the phenomenon can be explained by this parameter. The real
part of this parameter determines the decay the entanglement and
the imaginary part of it determines the oscillations of the
revival. We also have found that whenever $\gamma(t)$ takes
negative values, $E_a$ will begins to revive, and the
corresponding $E_r$ begins to decrease; when $\gamma(t)$ takes
positive values, $E_a$ will begin to decay, and $E_r$ will begin
to increase. We have also found that for this decoherence channel
the cluster state is the most robust to decoherence comparing with
Dicke, GHZ, and W states.

\section*{Acknowledgment}
This work was supported by National Natural Science Foundation of
China (grant no. 10974016).

\newpage

\begin{center}
\large\bf CAPTIONS \normalsize\rm
\end{center}

Figure 1: In the Markovian regime ($\lambda=10\gamma_0$,
$\delta=0$) $E_a$ and $E_r$ as functions of scaled time
$\gamma_0t$ for various initial states: (a) the atomic subsystem;
(b) the reservoir subsystem.

Figure 2: In the non-Markovian regime with relatively short
reservoir correlation time ($\lambda=0.1\gamma_0$, $\delta=0$)
$E_a$ and $E_r$  as functions of scaled time $\gamma_0t$ for
various initial states: (a) the atomic subsystem; (b) the
reservoir subsystem.

Figure 3: In the non-Markovian regime with relatively long
reservoir correlation time ($\lambda=0.01\gamma_0$, $\delta=0$)
$E_a$ and $E_r$  as functions of scaled time $\gamma_0t$ for
various initial states: (a) the atomic subsystem; (b) the
reservoir subsystem.

Figure 4: In the non-Markovian regime with relatively short
reservoir correlation time with detuning ($\lambda=0.1\gamma_0$,
$\delta=8\lambda$) $E_a$ and $E_r$  as functions of scaled time
$\gamma_0t$ for various initial states: (a) the atomic subsystem;
(b) the reservoir subsystem.

Figure 5: In the non-Markovian regime with relatively long
reservoir correlation time with detuning ($\lambda=0.01\gamma_0$,
$\delta=8\lambda$) $E_a$ and $E_r$  as functions of scaled time
$\gamma_0t$ for various initial states: (a) the atomic subsystem;
(b) the reservoir subsystem.

Figure 6: The atomic entanglement $E_a$, reservoir entanglement
$E_r$ and the decay rate $\gamma(t)$ for initial W state as
functions of $\gamma_0t$ ($\lambda=0.01\gamma_0$ and $\delta=0$):
the atomic subsystem (solid line); the reservoir subsystem (dashed
line); the decay rate (dotted line).

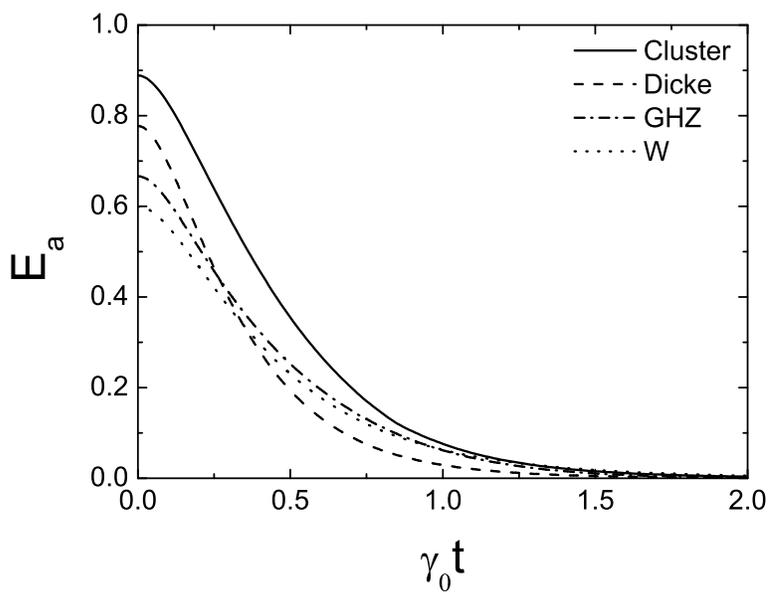
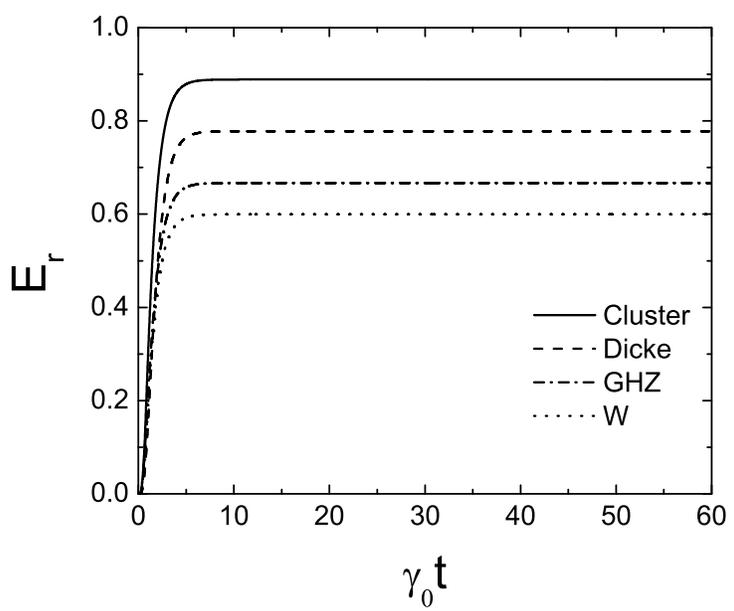
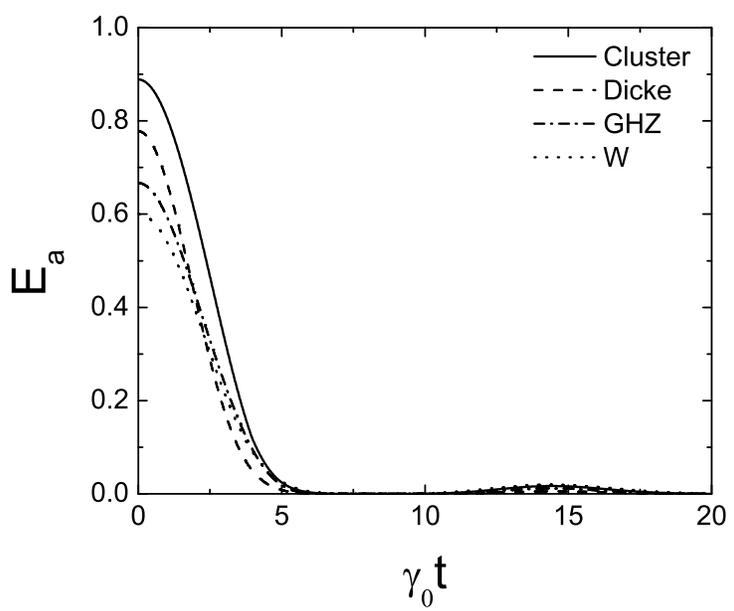
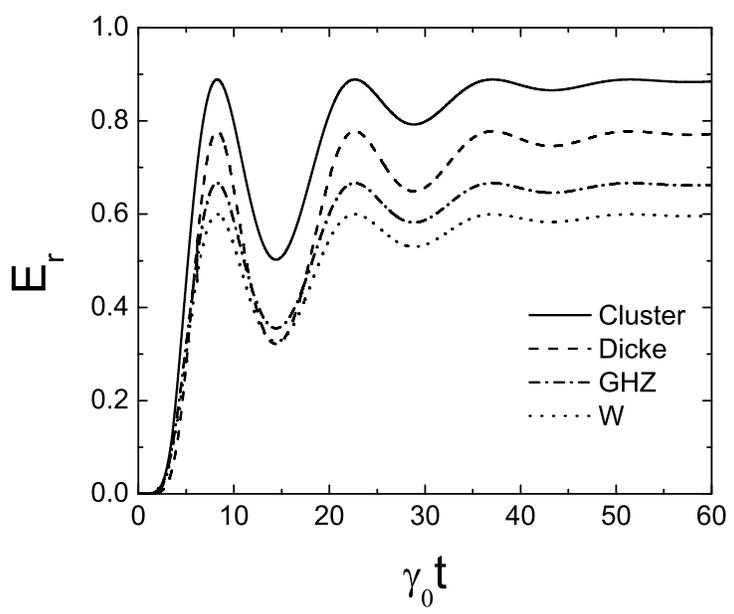
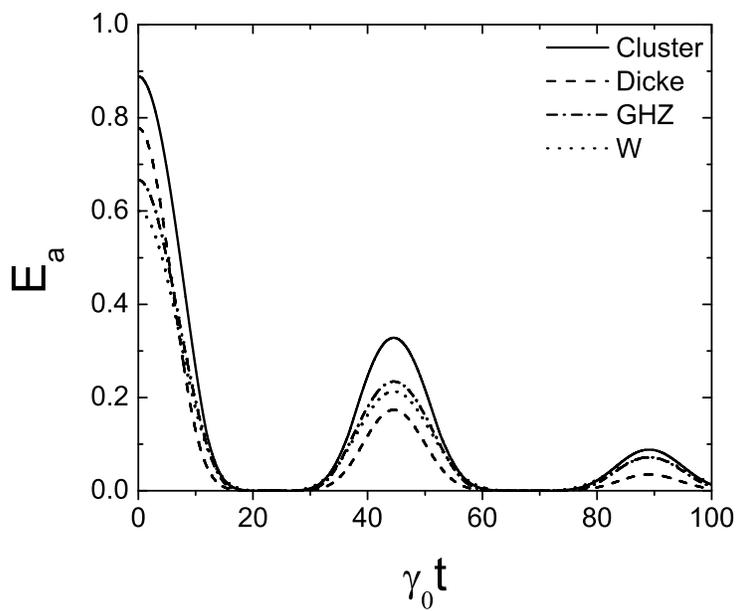
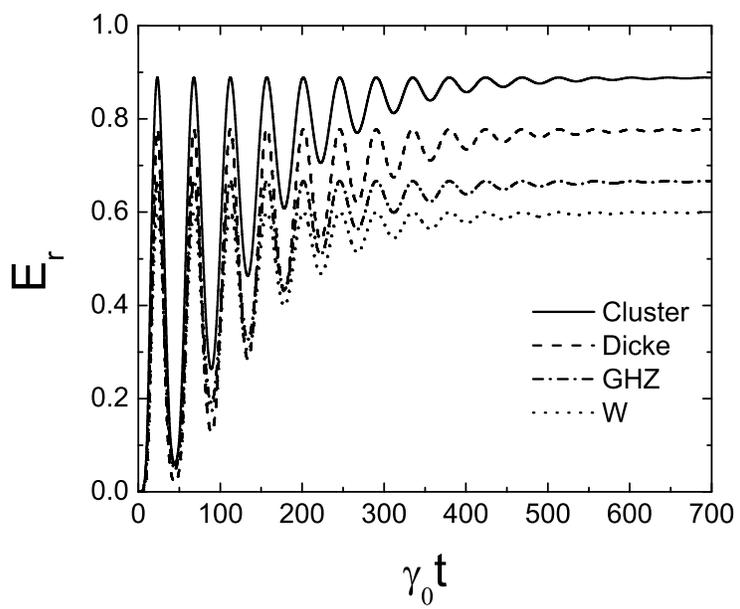
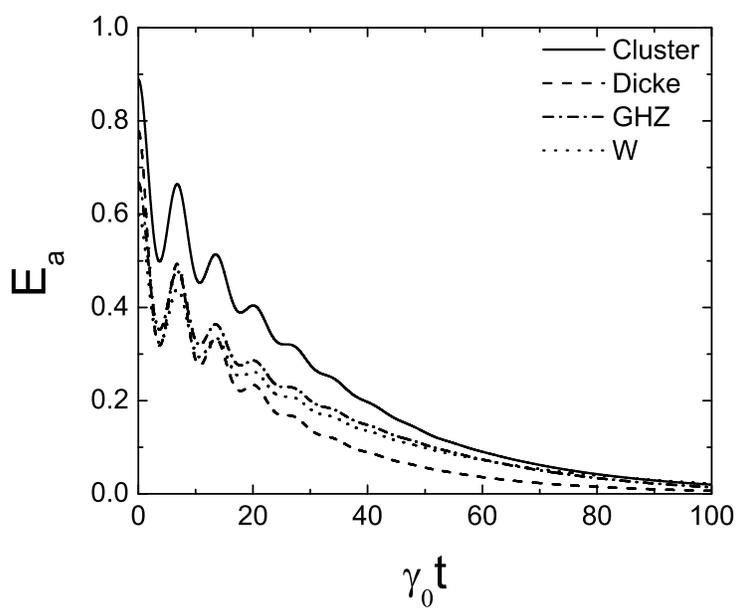
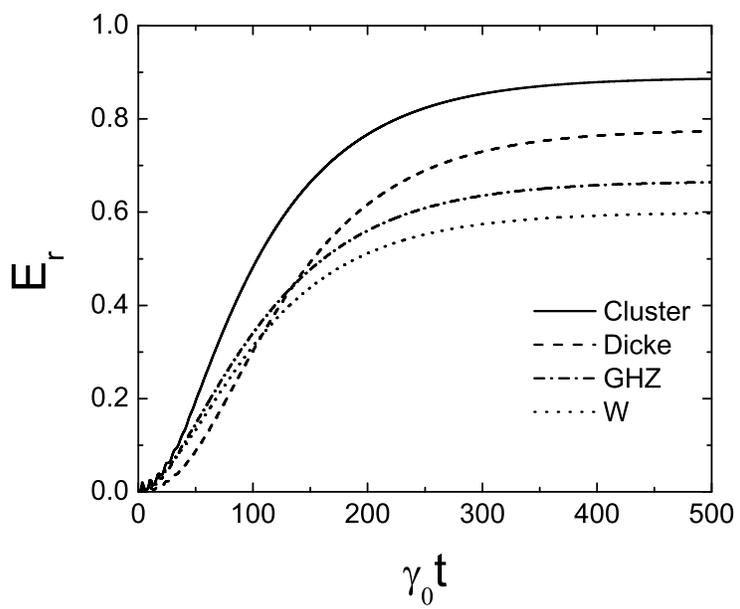
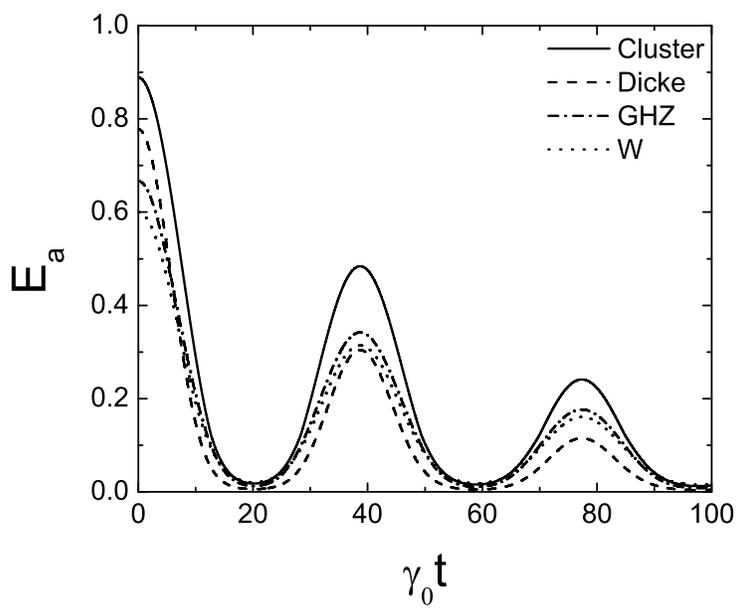
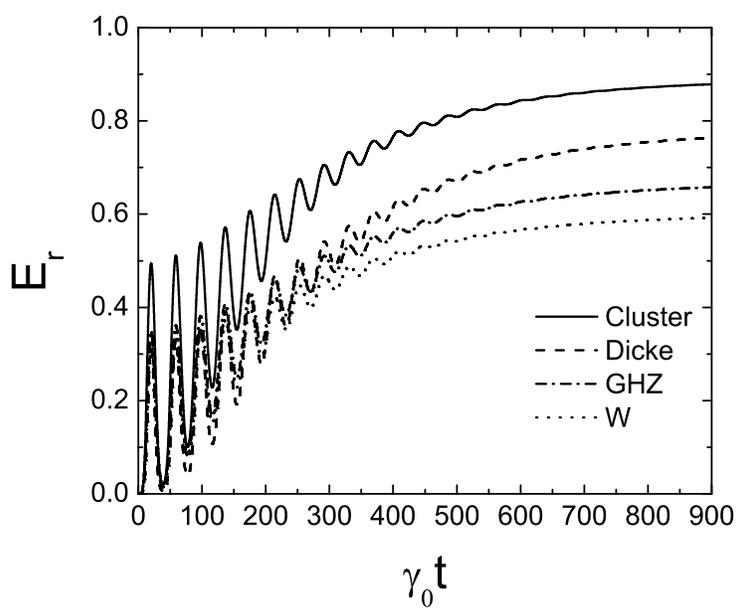
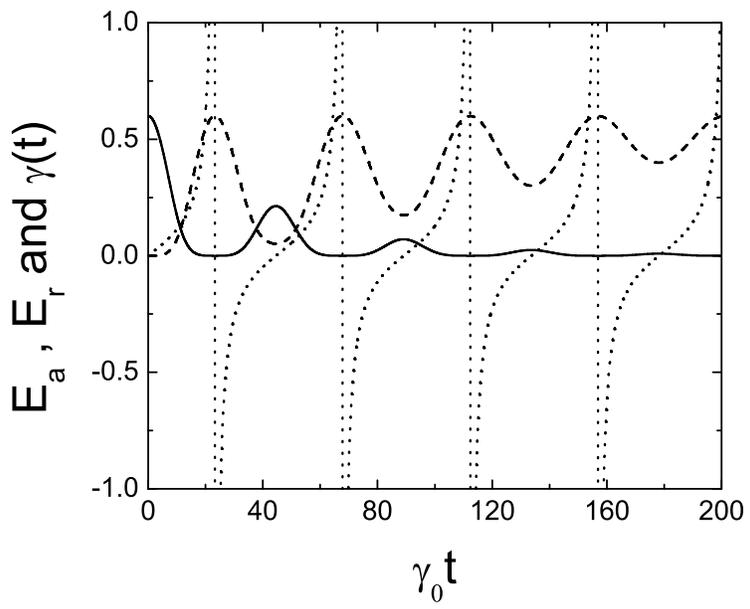
\end{document}